\documentclass{llncs}

\usepackage{xcolor}
\usepackage{graphicx}
\usepackage{booktabs}
\usepackage{multirow}
\usepackage{float}
\usepackage{listings}

\definecolor{grey}{rgb}{0.9,0.9,0.9}

\title{SEPIA: Search for Proofs Using Inferred Automata\thanks{The final publication is available at http://link.springer.com.}}

\author{Thomas Gransden, Neil Walkinshaw and Rajeev Raman}
\institute{Department of Computer Science, University of Leicester, UK.\\ \email{tg75@student.le.ac.uk, {\{nw91,rr29\}}@leicester.ac.uk}}

\begin{document}
\raggedbottom
\maketitle

\begin{abstract}
This paper describes SEPIA, a tool for automated proof generation in Coq. SEPIA combines model inference with interactive theorem proving. Existing proof corpora are modelled using state-based models inferred from tactic sequences. These can then be traversed automatically to identify proofs. The SEPIA system is described and its performance evaluated on three Coq datasets. Our results show that SEPIA provides a useful complement to existing automated tactics in Coq.
\end{abstract}

\section{Introduction}
Interactive theorem provers (ITPs) such as Coq \cite{Coq:manual} and Isabelle \cite{Isabelle02} are systems that enable the manual development of proofs for a variety of domains. These range from mathematics through to complex software and hardware verification. Thanks to the expressive logics that are used, they provide a very rich programming environment.

Nevertheless, constructing proofs can be a challenging and time-consuming process. A proof development will typically contain many routine lemmas, as well as more complex ones. The ITP system will take care of the bookkeeping and perform simple reasoning steps; however much time is spent manually entering the requisite tactics (even for the most trivial lemmas). In 2008, Wiedijk stated that it takes up to one week to formalize a page of an undergraduate mathematics textbook \cite{Freek08}.

To help combat this problem, we present SEPIA (Search for Proofs Using Inferred Automata) -- an automated approach designed to assist users of Coq. SEPIA automatically generates proofs by inferring state-based models from previously compiled libraries of successful proofs, and using the inferred models as a basis for automated proof search.

\section{Background}
\label{sec:background}
This section presents the necessary background required for this paper. We briefly introduce the underlying model inference technique (called MINT), followed by a motivating example.

\subsection{Inferring EFSMs with MINT}
\label{sub:inferringEfsms}
MINT \cite{WalkinshawWCRE} is an technique designed to infer state machine models from sequences, where the sequencing of events may depend on some underlying data state. Such systems are modelled as extended finite state machines (see Definition \ref{def:efsm}). EFSMs can be conceptually thought of as conventional finite state machines with an added memory. The transitions in an EFSM not only contain a label, but may also contain guards that must hold with respect to variables contained in the memory.

\begin{definition}\textbf{\emph{Extended Finite State Machine}}
\label{def:efsm}
 An Extended Finite State Machine (EFSM) $M$ is a tuple $(S,s_0,F,L,V,\Delta,T)$. $S$ is a set of states, $s_0 \in S$ is the initial state, and $F \subseteq S$ is the set of final states. $L$ is defined as the set of labels. $V$ represents the set of data states, where a single instance $v$ represents a set of concrete variable assignments. $\Delta:V \rightarrow \{True,False\}$ is the set of \emph{data guards}. Transitions $t \in T$ take the form $(a,l,\delta,b)$, where $a,b \in S$, $l \in L$, and $\delta \in \Delta$. 
\end{definition}

MINT infers EFSMs from sets of \emph{traces}. These can be defined formally as follows:

\begin{definition}
\label{def:traces}
A \emph{trace} $T=\langle e_0,\ldots,e_n\rangle$ is a sequence of $n$ trace elements. Each element $e$ is a tuple $(l,v)$, where $l$ is a label representing the names of function calls or input / output events, and $v$ is a string containing the parameters (this may be empty).
\end{definition}

The inference approach adopted by MINT \cite{WalkinshawWCRE} is an extension of a traditional state-merging approach \cite{Lang1998} that has been proven to be successful for conventional (non-extended) finite state machines \cite{WalkinshawStamina}. Briefly, the model inference starts by arranging the traces into a \emph{prefix-tree}, a tree-shaped state machine that exactly represents the set of given traces. The inference then proceeds by a process of \emph{state-merging}; pairs of states in the tree that are roughly deemed to be equivalent (based on their outgoing sequences) are merged. This merging process yields an EFSM that can accept a broader range of sequences than the initial given set.

The transitions in an EFSM not only imply the sequence in which events can occur, but also place constraints on which parameters are valid. This is done by inferring data-classifiers from the training data -- each data guard takes the following form $(l,v,possible)$ where $l \in L$, $v \in V$ and $possible \in$ \{$true, false$\}. When states are merged, the resulting machine is checked to make sure it remains consistent with the data guards.

\subsection{Motivating Example}
To motivate this work, we consider a typical scenario that arises during interactive proof. Suppose that we are trying to prove the following conjecture: \texttt{forall n m p:nat, p + n <= p + m -> n <= m}. The automated Coq tactics \cite{BC04} have only been able to perform routine reasoning (namely calling the \texttt{intros} tactic) to advance the proof to the following:

\begin{lstlisting}
  n : nat
  m : nat
  p : nat
  H : p + n <= p + m
  ============================
  n <= m
\end{lstlisting}
 
%\begin{minted}[bgcolor=bg]{text}
%  n : nat
 % m : nat
 % p : nat
  %H : p + n <= p + m
  %============================
   %n <= m
%\end{minted}

There are 2 theories from the Coq Standard Library called \texttt{Le.v} and \texttt{Lt.v}, that contain proofs about similar properties. The built-in tactics fail to prove the goal. The question we are faced with is this: Given the examples of successful proofs, can we use these to automatically find a proof for the above conjecture? 

In previous work \cite{GransdenCICM} we showed how to use MINT to infer EFSM models of Coq proofs. The resulting EFSMs were simply presented and used manually to derive proofs. This work extends our previous approach by automating the search process, allowing proofs to be completed automatically.

\section{SEPIA System Description}
\label{sec:implementation}
\begin{figure}[t]
	\centering
	\includegraphics[scale=0.45]{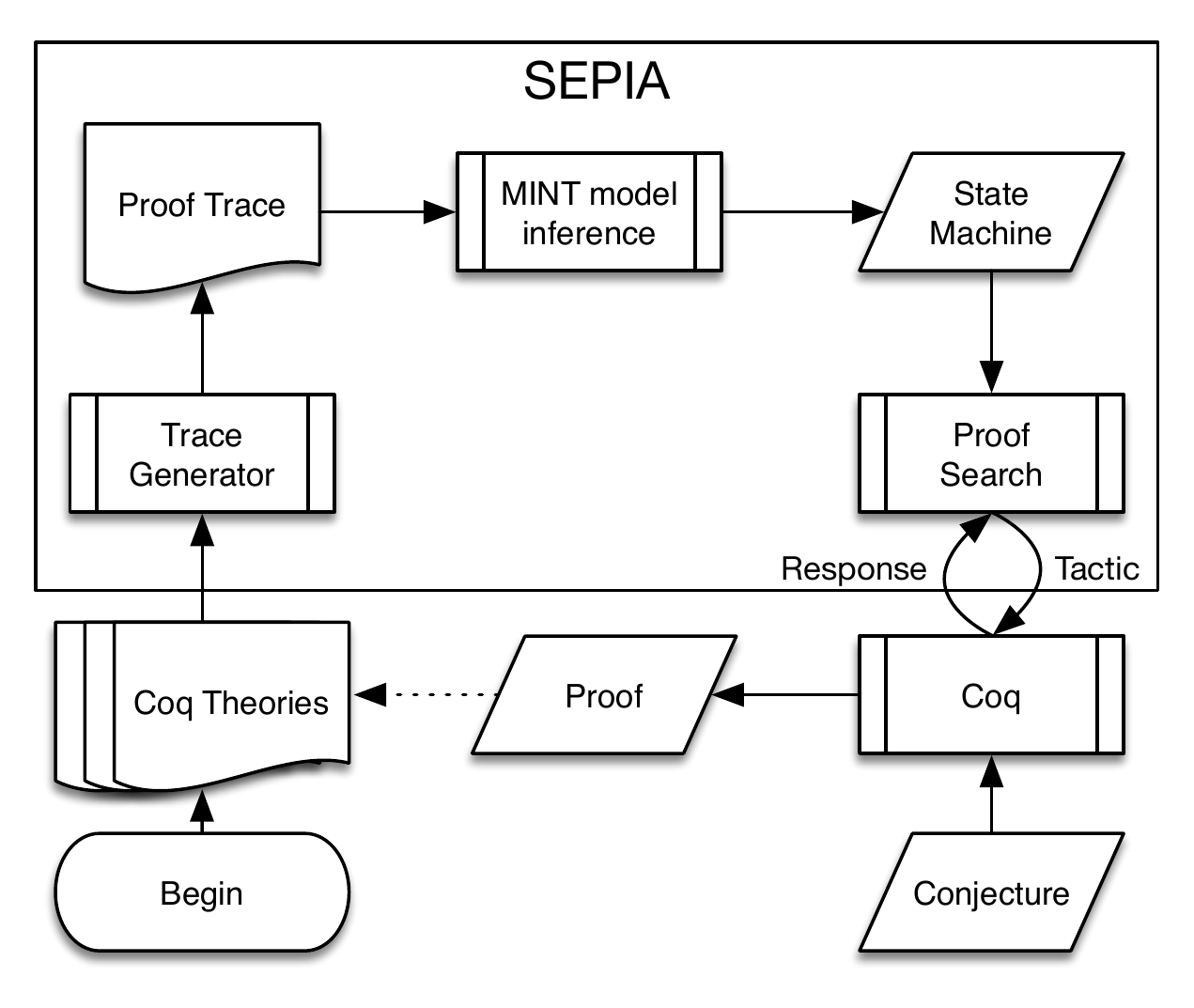}
	\caption{SEPIA overview}
	\label{fig:system}
\end{figure}

In this section we describe the SEPIA approach. We present the key stages of the technique. It is available\footnote{https://bitbucket.org/tomgransden/efsminferencetool} as a ProofGeneral extension that works with Coq. An overview of SEPIA is shown in Figure \ref{fig:system}. It contains three main stages:
\begin{enumerate}
\item{Generate proof traces from a selection of existing Coq theories.}
\item{Use MINT to infer a model from these proof traces.}
\item{Systematically search the model, formulating and attempting possible proofs from paths through the model.}
\end{enumerate}

Before describing these three steps in more detail, we look at three properties of the approach that are particularly appealing:

\paragraph{Adaptivity} 
For every iteration, as more valid proofs are discovered they can be incorporated into future cycles to infer more accurate models, forming a `virtuous loop'. This is a major benefit over the existing built-in automated tactics, which are typically limited to attempting a fixed set of tactics.

\paragraph{Automation}
Aside from providing the initial set of theories from which to infer a model, the user is not prompted for any other input. In addition, as will be elaborated later, the overall process typically completes in less than a minute (at least in the context of our experiments).

\paragraph{Ability to identify new proofs}
The state-merging process \cite{WalkinshawWCRE} can result in models that accept sequences of tactics which aren't present in the initial set of proofs. These wouldn't necessarily be intuitive, or be spotted from manual scrutiny of the proof library. These can however contain valuable steps that lead to a successful proof.

\subsection{Generating traces from existing proofs}
To begin a proof attempt we must provide one or more Coq theories from which we wish to generate a model. The proofs within the theories must  be converted into their corresponding \emph{proof traces} (see Definition \ref{def:traces}). This step is identical to the process used in our previous work \cite{GransdenCICM}.

Figure \ref{fig:traces} shows the proof script from the lemma \texttt{le\_antisym} from \texttt{Le.v} and the corresponding proof trace. An important concept in Coq proofs is the semicolon operator. If two (or more) tactics are separated by a semicolon, for example \texttt{t1;t2}, this means apply \texttt{t1} to the current goal and then apply \texttt{t2} to \emph{all} generated subgoals. We record the usage of the semicolon in our traces, so that this information can be reused during proof search.
\begin{figure}
\centering
\begin{tabular}{ c  c }
(a) Proof Script & (b) Trace\\
&\\
\begin{minipage}{0.35\textwidth}
\begin{lstlisting}
intros n m H; 
destruct H as [|m' H]; 
auto with arith.
intros H1.
absurd (S m' <= m'); 
auto with arith.
apply le_trans with n; 
auto with arith.
\end{lstlisting}
\end{minipage} & \begin{minipage}{0.75\textwidth}
\centering
\begin{tabular}{l|l|l}
Event $e$ & Label $l$ & Params $v$\\
\hline
$e_0$ &intros&``n m H;"\\
$e_1$ &destruct&``H as [$|$m' H];"\\
$e_2$ &auto&``with arith"\\
$e_3$ &intros&``H1"\\
$e_4$ &absurd&``(S m' $<=$ m');"\\
$e_5$ &auto&``with arith"\\
$e_6$ &apply&``le\_trans with n;"\\
$e_7$ &auto&``with arith"
\end{tabular}
\end{minipage} \\
\end{tabular}
\caption{Original proof and proof trace for an example lemma}
\label{fig:traces}
\end{figure}

\subsection{Inferring the model}
Once the proof traces have been generated, MINT is invoked to infer a model. There are two main parameters associated with MINT. The inference strategy dictates how states are merged during the inference process. A value called $k$ represents the minimum score before a pair of states can be deemed to be equivalent. An in-depth discussion of these variables is outside the scope of this paper. 

\begin{figure}[H]
\centering
\includegraphics[scale=0.25]{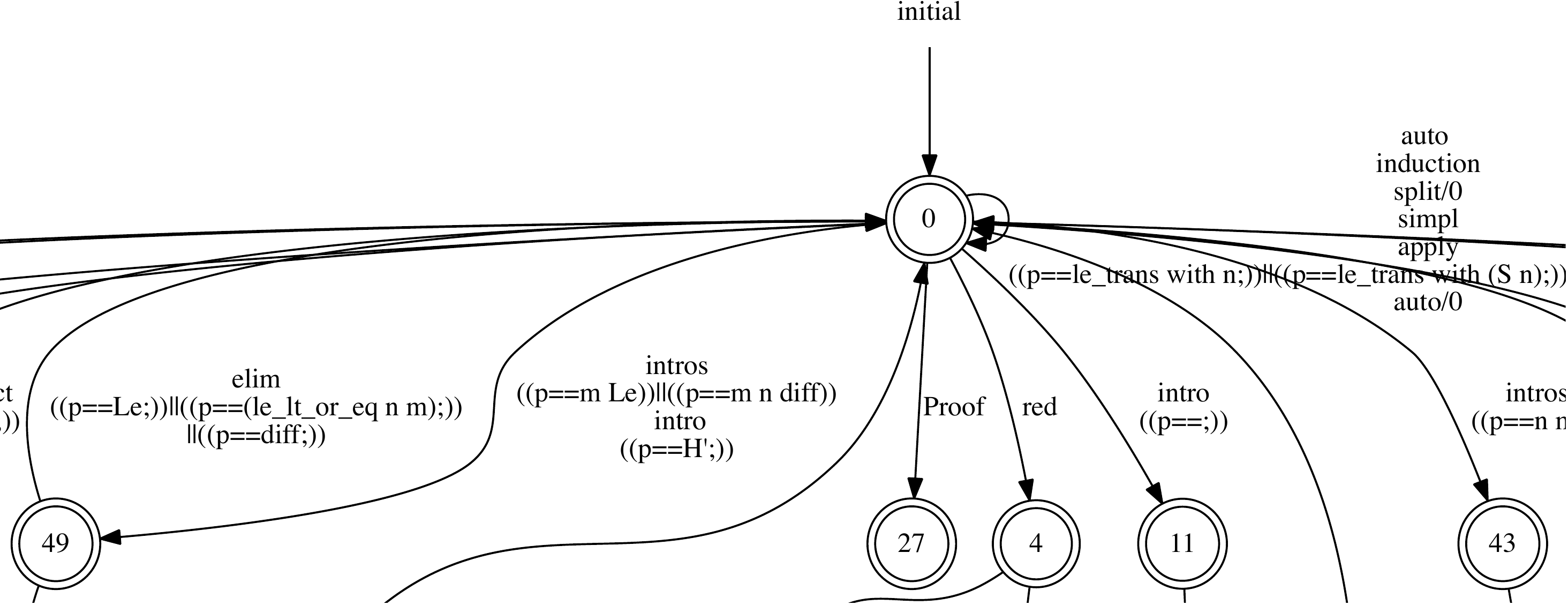}
\caption{Portion of inferred EFSM from {\tt Le.v} and {\tt Lt.v}}
\label{fig:efsmexample}
\end{figure}

A preliminary study (with results online) found that using the state merging strategy \texttt{redblue} and $k=1$ performed reasonably well for the task of interactive proving. These settings are based on the number of proofs discovered, the time taken and the presence of shorter/novel proofs. For the rest of this paper we refer to these as the default settings for MINT. A portion of the EFSM inferred from {\tt Le.v} and {\tt Lt.v} is shown in Figure \ref{fig:efsmexample}.

\subsection{Searching for a proof}
Once a model has been inferred it can be used to search for candidate proofs. We adopt a breadth-first search as this ensures that if a proof is contained in the model, the shortest one will be returned. An instance of Coq is loaded, and the lemma is stated. The proof search moves through the model and applies the tactics and arguments suggested on each transition. 

A timeout or a limit on the number of tactics applied can be provided to control the search. If we reach a point where a proof is found, SEPIA outputs the proof (and some proof search statistics). When running SEPIA on our motivating example we obtain the following result:

\begin{lstlisting}
Proof was: intros m n diff. elim diff; auto with arith.
5611 tactics evaluated.
Inference and search took 0 min, 1 sec
\end{lstlisting}

The above proof is particularly interesting for two reasons. Firstly, we have managed to prove something completely automatically that Coq's automated tools could not. Secondly, the sequence of tactics (and parameters) was not found anywhere else within \texttt{Le.v} or \texttt{Lt.v}.

\section{Evaluation}
\label{sec:eval}
In this section we provide an experimental evaluation of our approach. We consider the following research questions:

\begin{itemize}
\item{{\bf RQ1: }Can proofs be derived automatically using our approach?}
\begin{itemize}
    \item (a): How many proofs can be found?
    \item (b): How long does it take to find a proof?
  \end{itemize}
\item{{\bf RQ2: }Are there ``interesting" characteristics of the proofs?}
\begin{itemize}
    \item (a): Do the proofs contain new sequences of tactics?
    \item (b): Are the proofs shorter?
  \end{itemize}
\item{{\bf RQ3: }How does our results compare to Coq's built-in automated tactics?}
\end{itemize}

\subsection{Methodology}
The aim of this evaluation is to assess the practicalities of using our approach in real proof developments. We evaluate SEPIA on three distinct Coq contributions as our datasets. We use a method inspired by $k$-folds cross-validation \cite{KohaviIJCAI} in order to study proof attempts made by our approach.

\vspace{-3mm}

\subsubsection{Datasets}
The datasets used in this evaluation consist of theories selected from three Coq proof developments. The datasets were chosen mainly for their domain, complexity and size. All theories were selected before the experiments took place. SSreflect\footnote{http://ssr.msr-inria.inria.fr/doc/ssreflect-1.4/} contains seven core theories. We select all of these theories as our first dataset. Secondly, MSets\footnote{https://coq.inria.fr/library/} is an implementation of finite sets using lists/trees. All eleven theories are selected to form our second dataset. Finally, we use some theories from CompCert\footnote{http://compcert.inria.fr/doc/index.html}. Owing to the size of the development, we select a four theories containing both general purpose proofs along with some more specialized ones. Due to the exploratory nature of this evaluation, there are some threats to validity associated with the selection of data. We have only used three Coq datasets, so any results cannot be interpreted to represent performance on all Coq proofs.

\vspace{-3mm}

\subsubsection{Evaluating Proof Attempts}
To provide some answers to RQ1, we want to model the following situation: given some existing proofs, can we use these to prove new properties that are not part of the initial collection. To do this, we use an approach inspired by \emph{k}-folds cross-validation \cite{KohaviIJCAI}.

Each Coq theory file is taken individually and the proofs are randomly partitioned into $k$ non-overlapping sets. We then infer a model from $k-1$ of the sets, and try and prove the lemmas in the remaining set. This process repeats until each set has been used exactly once as the collection of lemmas to be proved.

For each proof attempt, we allow 10,000 tactics to be applied before reporting a failure. The results presented in this paper are from using $k=10$, a standard value for $k$-folds cross-validation \cite{KohaviIJCAI}. Other values of $k$ have been investigated and the full set of results are online.

As well as capturing whether a proof attempt was successful or not, when a proof is found we analyse how ``interesting" the proof is. First, we check and see whether a proof is shorter than the corresponding hand-curated proof. We also check whether the sequence of tactics was new (i.e. not present in the examples the model was inferred from). These provide us with answers to RQ2.

To investigate RQ3 we also run the Coq automated tools to try and prove each lemma. The following command is issued to Coq: \texttt{auto with * || eauto with * || tauto || firstorder || trivial}. This simply attempts to prove a goal by trying all of the automated tactics. The default search depth is used in all cases. Where we can specify lemma databases, we allow any available database to be used during proof search.

\subsection{Results}
The full results from our experiments are shown in Table \ref{tab:res}. The results are presented for each theory, grouped by library. The remainder of this section provides some answers to the research questions defined earlier.

\begin{table}[t]
\centering
\caption{Results Summary}
\label{tab:res}
\def\arraystretch{1.2}
\setlength\tabcolsep{1.5mm}
\begin{tabular}{llrrrrr} 
&&&\multicolumn{3}{c}{\textbf{SEPIA}}&\\
\cmidrule(r){4-6}
\textbf{Library}&\textbf{Theory}&\textbf{Size}&\textbf{Total}&\textbf{New}&\textbf{Shorter}&\textbf{Coq-Tacs}\\ 
\midrule 
\multirow{7}{*}{SSreflect}&ssrnat&341&135 (39\%)&14&9&59 (17\%)\\
&ssrbool&240&120 (50\%)&17&10&60 (25\%)\\
&seq&394&94 (24\%)&14&6&18 (4\%)\\
&fintype&243&42 (17\%)&15&1&0 (0\%)\\
&eqtype&82&36 (44\%)&18&2&10 (12\%)\\
&choice&30&6 (20\%)&0&0&1 (3\%)\\
&ssrfun&30&5 (16\%)&1&0&7 (23\%)\\
\midrule
\multirow{11}{*}{MSets}&avl&26&0 (0\%)&0&0&0 (0\%)\\
&decide&22&18 (81\%)&0&3&4 (18\%)\\
&eqproperties&106&43 (40\%)&1&5&47 (44\%)\\
&facts&65&17 (26\%)&4&8&10 (15\%)\\
&gentree&61&9 (15\%)&3&3&3 (5\%)\\
&list&42&8 (19\%)&3&3&3 (7\%)\\
&positive&67&13 (19\%)&5&4&1 (1\%)\\
&properties&137&78 (57\%)&9&3&15 (11\%)\\
&rbt&89&12 (13\%)&10&6&2 (2\%)\\
&tofiniteset&14&5 (35\%)&2&2&4 (28\%)\\
&weaklist&27&8 (30\%)&4&5&6 (22\%)\\
\midrule
\multirow{4}{*}{CompCert}&cshmgenproof&65&15 (23\%)&14&14&0 (0\%)\\
&amsgenproof0&57&12 (21\%)&9&9&6 (10\%)\\
&coqlib&114&36 (31\%)&24&23&16 (14\%)\\
&values&99&20 (20\%)&17&13&5 (5\%)\\
\bottomrule 
\end{tabular}
\end{table}

\vspace{-3mm}

\subsubsection{RQ1(a): A significant proportion of the lemmas were proved automatically using our approach}
In Table \ref{tab:res}, the column headed SEPIA shows the total number of lemmas proved in each theory using our approach. The results suggest that EFSM-based methods are useful at finding proofs automatically. Looking at each dataset as a whole, 32\% (438 out of 1360) of the SSreflect dataset were proved. In MSets, 30\% (211 out of 687) were successfully proved using our approach. In our selection of CompCert theories, there were 25\% (83 out of 335) proved. 

\vspace{-3mm}

\subsubsection{RQ1(b): Many proofs were discovered in under 30 seconds}
We measured the time required to derive a proof using our approach. These times take into account both the time required to infer the model and the search time. Over 90\% of the proofs were found within 30 seconds. These results show that when a user invokes the process, a proof will usually be delivered quickly. Overall, a proof can be discovered in a relatively small period of time. Of course, this is encouraging for the user involved in the proof development.

\vspace{-3mm}

\subsubsection{RQ2(a): A quarter of the proofs found were new sequences of tactics}
The number of new proofs discovered using our approach are listed under the `New' column in Table \ref{tab:res}. We compare the discovered proof with the ones used to infer the model If the sequence is not contained in an existing proof, then it is considered new and only found as a result of inferring an EFSM. Our results show a significant number of new proofs were discovered, backing up further that EFSMs can be useful for automated proof generation. In SSreflect, a total of 79 proofs were new. In the MSets theories, 41 new proofs were found, and 64 were discovered in CompCert.

\vspace{-3mm}

\subsubsection{RQ2(b): Many proofs discovered were shorter than their original ones}
We have listed the number of shorter proofs found in Table \ref{tab:res} under the Shorter column. When a proof is found, we compare the discovered proof with the original hand-curated one. The length (in terms of tactics used) of both proofs are then compared, to see if we managed to derive a shorter one. In SSreflect, 28 of the proofs found were shorter than their original counterparts. For MSets, 42 of the proofs were shorter, whilst in CompCert 59 of proofs were shorter. The combination of the state merging algorithms and a breadth-first search means we were able to identify these shorter proofs.

\vspace{-3mm}

\subsubsection{RQ3: SEPIA provides an alternative to existing Coq tactics}
The column headed Coq-Tacs in Table \ref{tab:res} provides the number of lemmas that were proved using Coq's automated tactics. Despite being relatively limited in the steps that they try, they manage to prove 155 SSreflect lemmas, 95 MSets lemmas and 27 of the CompCert lemmas. On the whole, we see that our approach significantly outperforms the automated tactics in terms of number of lemmas proved. This is to be expected, as they only provide modest automation. Nevertheless, there are occasions where the automated tactics prove more lemmas (in \texttt{msetproperties} and \texttt{ssrfun} for instance). 

\section{Related Work}
\label{sec:related}
There have been many projects aimed at improving the automation of proofs in ITPs. As we have shown in this work, machine learning can be applied in the context of interactive theorem proving. Specifically, we have shown that the tactics used in proofs can serve as useful features for machine learning algorithms. This is an area that has received moderate attention previously.

Jamnik \emph{et al.} have previously applied an Inductive Logic Programming technique to examples of proofs in the $\Omega$mega system \cite{Jamnik03}. Given a collection of well chosen proof method sequences, Jamnik \emph{et al.} perform a method of least generalisation to infer what are ultimately regular grammars. The value of even basic models is intuitive. Proofs could be derived automatically using the technique. However, the proof steps learned do not contain any parameters. The parameters required are reconstructed after running the learning technique.

Another approach that concentrated on Isabelle proofs was implemented by Duncan \cite{Duncan07}. Duncan's approach was to identify commonly occurring sequences of tactics from a given corpora. After eliciting these tactic sequences, evolutionary algorithms were used to automatically formulate new tactics. The evaluation showed that simple properties could be derived automatically using the technique; however the parameter information was left out of the learning approach. 

\section{Conclusion and Future Work}
\label{sec:conclusion}
This paper has presented SEPIA, an approach to automatically generate proofs in Coq. This has been achieved by applying model inference techniques to interactive proof scripts. We have shown that even learning from tactic sequences, which is admittedly a simplistic view of interactive proofs, can provide effective proof automation. It would be interesting to see what can be achieved by using more sophisticated views such as the proof goal view \cite{Grov12}.

The overall process is fully automated our evaluation shows SEPIA performs well on a range of proofs from three varied Coq datasets. It succeeds in proving a number of lemmas that were out of reach for Coq's automated tactics. Additionally, when SEPIA finds a proof it usually does so in seconds. 

As well as reusing existing proofs, SEPIA can construct proofs using new tactic sequences. These new sequences might not have been identified if manually analysing proof libraries. In our evaluation, we also identified a number of shorter proofs (by comparing the proofs found using SEPIA to original proofs). This follows the trend of other comparisons of automated and human proofs \cite{Alama12}.

We plan to investigate automatic identification of appropriate theories or lemmas that could be used to infer models. Currently, we use whole theories; however it may be the case that only a handful of these proofs are actually useful. By using methods such as ML4PG \cite{ML4PG13} it may be possible to discover the most useful lemmas from a large collection of theories. 

%Bibliography
\bibliography{cade-bib}

\begin{thebibliography}{10}
\providecommand{\url}[1]{\texttt{#1}}
\providecommand{\urlprefix}{URL }

\bibitem{Alama12}
Alama, J., K{\"{u}}hlwein, D., Urban, J.: {Automated and Human Proofs in
  General Mathematics: An Initial Comparison}. In: Bj{\o}rner, N., Voronkov, A.
  (eds.) Logic for Programming, Artificial Intelligence, and Reasoning, LNCS,
  vol. 7180, pp. 37--45. Springer (2012)

\bibitem{BC04}
Bertot, Y., Cast\'eran, P.: Interactive Theorem Proving and Program Development
  - Coq'Art: The Calculus of Inductive Constructions. Springer (2004)

\bibitem{Duncan07}
Duncan, H.: The Use of Data Mining for the Automatic Formation of Tactics.
  Ph.D. thesis, University of Edinburgh (2007)

\bibitem{GransdenCICM}
Gransden, T., Walkinshaw, N., Raman, R.: {Mining State-Based Models from Proof
  Corpora}. In: Watt, S.M., Davenport, J.H., Sexton, A.P., Sojka, P., Urban, J.
  (eds.) Intelligent Computer Mathematics, LNCS, vol. 8543, pp. 282--297.
  Springer (2014)

\bibitem{Grov12}
Grov, G., Komendantskata, E., Bundy, A.: {A Statistical Relational Learning
  Challenge – Extracting Proof Strategies from Exemplar Proofs}. In: ICML-12
  Workshop on Statistical Relational Learning (2012)

\bibitem{Jamnik03}
Jamnik, M., Kerber, M., Pollet, M., Benzm{\"u}ller, C.: {Automatic Learning of
  Proof Methods in Proof Planning}. Logic Journal of the IGPL  11(6),  647--673
  (2003)

\bibitem{KohaviIJCAI}
Kohavi, R.: {A Study of Cross-validation and Bootstrap for Accuracy Estimation
  and Model Selection}. In: Proceedings of the 14th International Joint
  Conference on Artificial Intelligence. pp. 1137--1143. Morgan Kaufmann (1995)

\bibitem{ML4PG13}
Komendantskaya, E., Heras, J., Grov, G.: {Machine Learning in Proof General:
  Interfacing Interfaces}. In: User Interfaces for Theorem Provers. EPTCS, vol.
  118, pp. 15--41 (2013)

\bibitem{Lang1998}
Lang, K.J., Pearlmutter, B.A., Price, R.A.: {Results of the {Abbadingo} {One}
  {DFA} learning competition and a new evidence-driven state merging
  algorithm}. In: Honavar, V., Slutzki, G. (eds.) Proceedings of the 4th
  International Colloquium on Grammatical Inference. vol. 1433, pp. 1--12.
  Springer (1998)

\bibitem{Coq:manual}
\mbox{The Coq Development Team}: {The Coq Proof Assistant Reference Manual.
  Version 8.4}. LogiCal Project, \url{http://coq.inria.fr/refman}

\bibitem{Isabelle02}
Nipkow, T., Paulson, L.C., Wenzel, M.: Isabelle/HOL --- A Proof Assistant for
  Higher-Order Logic, LNCS, vol. 2283. Springer (2002)

\bibitem{WalkinshawStamina}
Walkinshaw, N., Lambeau, B., Damas, C., Bogdanov, K., Dupont, P.: {STAMINA: A
  Competition to Encourage the Development and Assessment of Software Model
  Inference Techniques}. Empirical Software Engineering  18(4),  791--824
  (2013)

\bibitem{WalkinshawWCRE}
Walkinshaw, N., Taylor, R., Derrick, J.: {Inferring Extended Finite State
  Machine Models from Software Executions}. Empirical Software Engineering pp.
  1--43 (2015)

\bibitem{Freek08}
Wiedijk, F.: {Formal Proof -- Getting Started}. Notices of the AMS  55(11),
  1408--1414 (Dec 2008)

\end{thebibliography}
\bibliographystyle{splncs03}

\end{document}